\documentclass[journal=nalefd,manuscript=letter,layout=onecolumn]{achemso}

\usepackage{chemformula} 
\usepackage[T1]{fontenc} 
\usepackage[version=4]{mhchem}
\author{Nguyen Ha My Dang}
\affiliation{Universit\'{e} de Lyon, Institut des Nanotechnologies de Lyon, INL-UMR5270, CNRS, Ecole Centrale de Lyon, 36 avenue Guy de Collongue, Ecully, F-69134, France}
\author{Dario Gerace}
\affiliation{Dipartimento di Fisica, Universit\`{a} di Pavia, via Bassi 6, I-27100 Pavia, Italy}
\author{Emmanuel Drouard}
\affiliation{Universit\'{e} de Lyon, Institut des Nanotechnologies de Lyon, INL-UMR5270, CNRS, Ecole Centrale de Lyon, 36 avenue Guy de Collongue, Ecully, F-69134, France}
\author{Ga\"{e}lle Tripp\'e-Allard}
\affiliation{Laboratoire Aim\'e Cotton, CNRS, Univ. Paris-Sud, ENS Paris-Saclay, Universit\'e Paris-Saclay, 91405 Orsay Cedex, France}
\author{Ferdinand L\'ed\'ee}
\affiliation{Laboratoire Aim\'e Cotton, CNRS, Univ. Paris-Sud, ENS Paris-Saclay, Universit\'e Paris-Saclay, 91405 Orsay Cedex, France}
\author{Radoslaw Mazurczyk}
\affiliation{Universit\'{e} de Lyon, Institut des Nanotechnologies de Lyon, INL-UMR5270, CNRS, Ecole Centrale de Lyon, 36 avenue Guy de Collongue, Ecully, F-69134, France}
\author{Emmanuelle Deleporte}
\affiliation{Laboratoire Aim\'e Cotton, CNRS, Univ. Paris-Sud, ENS Paris-Saclay, Universit\'e Paris-Saclay, 91405 Orsay Cedex, France}
\author{Christian Seassal}
\affiliation{Universit\'{e} de Lyon, Institut des Nanotechnologies de Lyon, INL-UMR5270, CNRS, Ecole Centrale de Lyon, 36 avenue Guy de Collongue, Ecully, F-69134, France}
\author{Hai Son Nguyen}
\email{hai-son.nguyen@ec-lyon.fr}
\affiliation{Universit\'{e} de Lyon, Institut des Nanotechnologies de Lyon, INL-UMR5270, CNRS, Ecole Centrale de Lyon, 36 avenue Guy de Collongue, Ecully, F-69134, France}

\date{\today} 
\title
  {Tailoring dispersion of room temperature exciton-polaritons with perovskite-based subwavelength metasurfaces}
\abbreviations{HOP,PEPI,DMF,PMMA,PL,ARR,ARPL,UP,LP}
\begin{document}

\newpage 

\begin{abstract}
Exciton-polaritons, elementary excitations arising from the strong coupling regime between photons and excitons in insulators or semiconductors, represent a promising platform for studying quantum fluids of light and realizing prospective all-optical devices. Among different materials for room temperature polaritonic devices, two-dimensional (2D) layered perovskites have recently emerged as one of the promising candidates thanks to their prominent excitonic features at room temperature. Here we report on the experimental demonstration of exciton-polaritons at room temperature in resonant metasurfaces made from a subwavelength 2D lattice of perovskite pillars. These metasurfaces are obtained via spincoating, followed by crystallization of the perovskite solution in a pre-patterned glass backbone.  The strong coupling regime is revealed by both angular-resolved reflectivity and photoluminescence measurements, showing anticrossing between photonic modes and the exciton resonance with a Rabi splitting in the 200\,meV range.  Moreover, we show that the polaritonic dispersion can be engineered by tailoring the photonic Bloch mode to which perovskite excitons are coupled. Linear, parabolic, and multi-valley polaritonic dispersions are experimentally demonstrated. All of our results are perfectly reproduced by both numerical simulations based on a rigorous coupled wave analysis and an elementary model based on a quantum theory of radiation-matter interaction. Our results suggest a new approach to study exciton-polaritons and pave the way towards large-scale and low-cost integrated polaritonic devices operating at room temperature.
\end{abstract}

\newpage 
\section{Article}	

Exciton polaritons - half-light/half-matter elementary excitations arising from the strong coupling between excitons and photons~\cite{Weisbuch1992} offer unprecedented insight into fundamental physical phenomena, as well as exciting technological prospects. On the one hand, thanks to their photonic component, polaritons can exhibit ballistic propagation over macroscopic distances~\cite{Freixanet2000}. Their photonic nature also provides optical means to generate, probe, and detect polaritons. On the other hand, polaritons exhibit strong nonlinearities inherited from exciton-exciton interactions, which are orders of magnitudes more efficient than the  photon-photon nonlinear Kerr effect~\cite{PhysRevB.76.045319,PhysRevB.75.075332}. These unique features make polaritons an attractive playground to study fascinating physical phenomena such as out-of-equilibrium Bose-Einstein condensation~\cite{Kasprzak2006,Byrnes2014}, superfluidity~\cite{Amo2009}, quantum vortices~\cite{Lagoudakis2008}, analog gravity~\cite{Nguyen2015}, and topological insulators~\cite{St-Jean2017,Klembt2018}.  Furthermore, from an application point of view, these quasiparticles represent a promising platform to make all-optical devices such as polariton lasers~\cite{Kasprzak2006,Christopoulos2007,Daskalakis2013}, optical transistors~\cite{Ballarini2013a,Zasedatelev2019}, resonant tunneling diodes~\cite{Nguyen2013}, optical switches~\cite{Amo2010,DeGiorgi2012a}, and rooters~\cite{Marsault2015}.\\

Pioneering work on exciton-polaritons was performed with GaAs and CdTe based quantum wells embedded within vertical Fabry Perot cavities~\cite{Weisbuch1992}. Although these materials, especially GaAs, are still the most used excitonic materials for studying polaritonic physics, their operation is limited to cryogenic temperatures due to the modest excitonic binding energy of few meV~\cite{Sanvitto2016}. In view of making polaritonic devices suited for practical applications, a quest has been set towards room temperature excitons, and therefore the use of materials of high excitonic binding energy such as GaN~\cite{Christopoulos2007,Daskalakis2013}, ZnO~\cite{Zamfirescu2002,Franke_2012}, organic semiconductors~\cite{Lidzey1998,Takada2003,Holmes2004} and, more recently, monolayers of transition metal dichalcogenides (TMDCs)~\cite{Liu2014,Grosso2017}.\\

Other appealing candidates for room temperature polaritonic devices are hybrid organic-inorganic perovskites (HOPs). These materials possess remarkable optical properties mostly determined by their inorganic component, such as bandgap tunability, high luminescence quantum yield, and narrow emission line-width. They also exhibit key advantages of organic semiconductors as solution processability and the strong exciton binding energy needed for high temperature operation~\cite{Sutherland2016}. In particular, two-dimensional (2D) layered HOPs offer excitons with binding energies up to hundreds meV and superior oscillator strength in comparison with conventional inorganic quantum well excitons. Thanks to this robust binding energy, the strong coupling regime could be observed at room temperature. Indeed, 2D layered HOPs were employed as a highly relevant material for room temperature polariton even before the ``perovskite fever". Using these materials, observation of cavity-exciton polaritons and plasmon-exciton polaritons have already been reported by many groups~\cite{Fujita1998,Shang2018a,Lanty2009,Wei2012,Han2012,Nguyen2014,lanty2008strong,Wang2018,Fieramosca2018}. Most recently, it has been shown that the nonlinearity of HOP polaritons is mostly due to the exciton-exciton interaction~\cite{Fieramoscaeaav9967}. Such behavior, reserved so far only for GaAs polaritons at cryogenic temperature~\cite{Vladimirova2009}, suggests that 2D layered HOP would be the ideal material to exploit polariton physics at room temperature.\\

In this letter, we report on the experimental demonstration of exciton polaritons at room temperature in HOP-based subwavelength metasurfaces textured in a periodic 2D lattice. This is achieved when coupling the HOP excitons with the photonic Bloch modes of a lattice of HOP pillars. The strong coupling regime for different lattice designs and polarizations is observed in both angular-resolved reflectivity and photoluminescence measurements, showing Rabi splitting in the range of 200\,meV. Most interestingly, we show that the polaritonic dispersion can be engineered by tailoring the photonic Bloch mode to which HOP excitons are coupled. Polaritonic dispersion with linear, parabolic, and even multi-valley characteristics are observed. All of the experimental measurements are perfectly reproduced by both numerical simulations and a simplified quantum theory model, the latter suggesting that we can interpret these results in terms of elementary excitations previously defined as {\it photonic crystal polaritons}~\cite{Gerace2007,Bajoni2009}. Our results suggest a new approach to study exciton-polaritons beyond the textbook Fabry Perot configuration, and pave the way toward large-scale and low-cost integrated polaritonic devices operating at room temperature.\\

The active material employed in this work is the 2D layered HOP, namely bi-(phenethylammonium) tetraiodoplumbate, known as PEPI, with chemical formula (C$_6$H$_5$C$_2$H$_4$NH$_3$)$_2$PbI$_4$. Its molecular structure is described in Fig~1(a). The composition of alternating organic/inorganic monolayers features the multi-quantum well structure in PEPI:  the organic layers play the role of potential barriers that are able to confine the electronic excitation within the inorganic quantum wells~\cite{ISHIHARA1994,Fieramosca2018}. The confinement effect is strengthened thanks to the high dielectric contrast between the organic and inorganic layers~\cite{Hong1992}. The combination of quantum confinement and dielectric confinement results in the high exciton binding energy, ranging up to hundred meV at room temperature~\cite{Gauthron2010}. Additionally, PEPI has been reported to have higher nonlinearity related to the delocalized Wannier excitons in comparison with Frenkel excitons in all-organic materials~\cite{Abdel-Baki2016}. We first study the bare material:  50\,nm$-$thick PEPI film on flat SiO$_2$ substrate is prepared by spincoating, then immediately encapsulated  with Poly-methyl methacrylate (PMMA) on top to avoid contact with humidity.  X-ray diffraction (XRD) measurements [see Fig~1(b)] clearly evidence the critical peaks of PEPI crystalline structure, $[0\,0 \,2l]$, with $l=1-6$. The photoluminescence (PL) spectrum, under excitation at 3.06\,eV, exhibits a relatively sharp emission peak at 2.36\,eV with full width at half maximum of 0.062\,eV [see Fig~1(c)]. The absorption spectrum displays a strong absorption peak at 2.4\,eV, which is superimposed to the PL spectrum. These spectral features clearly show an excitonic behavior at room temperature. The absorption continuum in UV range corresponds to the absorption above the bandgap of the semiconductors. The strong excitonic peak and continuum observed in the absorption spectrum at room temperature are in good agreement with the results reported in the literature~\cite{Zhang2010}. \\

The fabrication process of PEPI metasurfaces is illustrated in Fig~1(d).  A periodically patterned SiO$_2$ backbone is first prepared by electron-beam lithography, and then undergoes ionic dry-etching with 150\,nm of etching depth. PEPI in DMF (dimethylformamide) solution is infiltrated inside the air holes of this backbone via spincoating, then crystallized with the help of thermal annealing to form a pillar-lattice of PEPI. The Scanning Electron Microscope (SEM) images confirm that almost the entire PEPI fills inside the holes after its deposition, even though there are still residues of PEPI outside. A 200\,nm thick layer of PMMA is immediately spincoated on top to encapsulate the whole structure. From the photonic point of view, the PMMA/PEPI nano-pillars/SiO$_2$ stack is equivalent to PEPI nano-pillars standing in a homogeneous optical medium, since the refractive index of PMMA (1.49) is closely matched with the SiO$_2$ one (1.47) in the whole spectral range of interest (400-600\,nm). At difference with previous reports on inorganic exciton polaritons at low temperature~\cite{Bajoni2009} in which the periodic metasurface was not directly modifying the active material, here the excitonic medium is strongly modulated by the periodic pattern.\\

The strong coupling regime in the PEPI metasurfaces is investigated by angle-resolved reflectivity (ARR) and angle-resolved PL (ARPL) measurements through a Fourier spectroscopy set up. These angular-resolved experiments are performed along $\Gamma$X direction of reciprocal space, using two distinct polarizations:the S-polarization, corresponding to non-zero $E_y$ electric field component, and the P-polarization, corresponding to non-zero $H_y$ magnetic field component [see Fig~1(e)].  Numerical simulations employing the Rigorous Coupled Wave Analysis (RCWA)\cite{Liu2012} are performed to predict the photonic Bloch mode response as a function of energy and wave vector, to be directly compared to the experimental results.\\

Two designs of PEPI metasurface are considered in our study [see table in Fig~1(e)]: 1/\underline{Structure A}: 80\,nm-height pillar-lattice with a d/a aspect ratio of 0.8 and a lattice parameter of 250\,nm; 2/\underline{Structure B:} 50\,nm-height pillar-lattice with a d/a aspect ratio of 0.9 and a lattice parameter of 350\,nm. These designs are chosen to provide a rich variety of photonic Bloch modes and mode dispersions in the vicinity of PEPI exciton energy. In fact, structure A only exhibits a S-polarized photonic mode in this spectral region, while structure B exhibits both S- and P-polarized modes. Their dispersions can be numerically predicted by simulating the ARR of passive structures, in which PEPI is replaced by a dielectric medium of refractive index $n=2.4$. The photonic dispersions of the two structures are strikingly different: i) Structure A displays a single S-polarized mode with an almost linear dispersion, as evidenced in Fig~2(a); ii) Structure B shows a P-polarized mode with an almost parabolic dispersion close to normal incidence [see Fig~2(b)], and three S-polarized modes [Fig~2(c)], two of which exhibit parabolic dispersions with opposite and small curvatures corresponding to a mini-gap opening, while the third one appears a quite exotic mode with a multi-valley dispersion of W-shape with two off-$\Gamma$ minima. \\

The coupling between the previous Bloch modes with PEPI excitons is first investigated by reflectivity measurements. Figures~2(d-f) present the ARR experimental results (right panels) directly compared to the numerical simulations (left panels) performed on the active structures. From these results, the strong coupling regime induced by the presence of the strong excitonic response is clearly evidenced: all dispersion curves are bent when approaching the exciton energy, and undergo the typical anticrossing behavior as a function of the in-plane wave vector (scanned through the incidence angle). The dispersion of each mode splits into an upper polariton (UP) and a lower polariton (LP) branch, respectively, with a very good overall agreement between measurements and numerical simulations.
Rabi splittings of approximately 200\,meV are measured. We note that the PEPI model used for numerical simulations only takes into account the excitonic resonance, but not the highly absorptive continuum states above the PEPI bandgap (see Methods section for more details). This explains the much weaker measured signal of UP as compared to the simulated one [Fig~2(d,f)]. In the case of P-polarized measurements of structure B [Fig~2(e)], the UP is not visible. This is also in agreement with the related simulations in the left panel, in which the simulated signal of this UP is already very weak.\\

Although the reflectivity experiment proves the existence of photonic crystal polaritonic modes in this PEPI metasurface, it is important to demonstrate that polariton states can be populated in these structures via optical pumping. Figures~2(g-i) present the ARPL experimental results (right panels) performed on the active structures. To get a qualitative comparison, simulations of angle-resolved absorption are also presented on equivalent color scale plots (left panels). Below the bare exciton, the LP emission is clearly observed in these PL measurements, confirming the existence of polariton states in our samples. Typical to other room temperature polaritonic systems using high bandgap materials, the UP is not observed in PL measurements~\cite{Lidzey1999,Plumhof2014}. Regarding the signal from a non-dispersive band corresponding to uncoupled PEPI excitons, which is evident both from PL measurements and absorption simulations with a small spectral shift corresponding to the Stokes-shift, we notice that depending on the location within the PEPI pillar, an exciton can be at an anti-node or a node position of the photonic mode, thus can undergo strong coupling or weak coupling with these modes. This is different from text-book exciton quantum well polaritons, where all the excitons are equally coupled to the same Fabry-Perot mode.\\

The strong coupling mechanism in periodically textured excitonic metasurfaces can be analyzed with a full quantum theory of radiation matter coupling taking into account the in-plane periodicity of photonic modes and excitonic wave functions on an equal footing. The full Hopfield matrix can be obtained from a classical solution of Maxwell equations in such periodically patterned multilayer structure, and solving for the corresponding Schr\"{o}dinger equation for the exciton envelope function, then coupling them to obtain the coupling matrix elements, as detailed in a previous work~\cite{Gerace2007}. Here, we approximate the excitonic response as if it was concentrated in a single quantum well layer at the center of the periodically patterned region, with an effective oscillator strength taking into account the finite thicknesss of the 2D HOP film. Polaritonic modes obtained by numerically diagonalizing the Hopfield matrix thus are presented in Figs~2(j-l). Correspondingly, the experimental data extracted from the ARR measurements are also superimposed, showing a remarkably good agreement over the whole parameters range.\\

We finally discuss on the engineering of polaritonic dispersion in these PEPI metasurfaces. The results from both ARR and ARPL measurements presented above show that the polariton dispersion shares the same shape as the one of uncoupled photonic modes when working out of the anticrossing region. Most remarkably, this leads to polaritonic modes that can be engineered to display linear [Figs~2(j)], slow-light [Figs~2(k)], or even  multi-valley [Figs~2(l)] characteristics. Notice that this is due to the strong periodic modulation of the active material, at difference with previous reports on photonic crystal modulated dispersions of guided polaritons with in-plane uniform active medium~\cite{Bajoni2009}. To quantitatively assess these different dispersions, we extracted the group velocity (normalized to the speed of light $c=3.10^8\,m/s$), $|v_g|/c$ from experimental data and calculations of polariton dispersions from Hopfield matrix diagonalization. In structure A, linear dispersion in LP corresponds to $|v_g|/c\sim 0.42$ when $|k_x|$ exceeds 5\,$\mu m^{-1}$. This linear polaritonic dispersion is similar to the one of guided polaritons recently reported from several groups~\cite{Lerario2017,Jamadi2018}. Such high velocity regime is perfectly adapted to study the ballistic propagation of polaritons for transmitting information and gating signals between polaritonic devices~\cite{Franke_2012}.  Figure~3(a) presents the normalized $|v_g|/c$ corresponding to P-polarized LP in the structure B. Contrary to the previous case, here polaritons undergo a slow-light regime with a maximum of $|v_g|/c\sim 0.08$ corresponding to the inflexion points of the dispersion at $k_x\approx\pm5\,\mu m^{-1}$. We further notice that this behavior is quite similar to microcavity polariton dispersion. Such slow-light regime, exhibiting high density of states, would be well suited to study polaritonic non-linear effects, as well as to trigger the Bose-Einstein condensation of polaritons. Finally, the group velocity corresponding to the S-polarized multi-valley band in the structure B is reported in Fig~3(b). The two valleys correspond to the vanishing of group velocity at $k_x\approx\pm1.5\,\mu m^{-1}$. To the best of our knowledge, this is the first experimental observation of multi-valley polaritonic dispersion. It could be the building block of polariton valleytronic physics~\cite{Sun2017,Karpov2018}. Indeed, recent theoretical works have predicted that Bose-Einstein condensation of polaritons with W-shaped dispersion would take place at the valley extrema~\cite{Karpov2018}, thus paving the way to study Josephson oscillation in momentum space~\cite{Zheng2018}, spontaneous symmetry breaking and two-mode squeezing~\cite{Sun2017}. Moreover, when not working at the extrema but at $k_x=0$, the W-shaped dispersion is also a perfect test bed for parametric scattering experiment~\cite{Baumberg2000}.\\

In conclusion, we propose periodically patterned excitonic metasurfaces as a novel platform to study exciton-polariton physics. In constrast to the traditional microcavity design, the metasurface approach offers high flexibility for the tailoring of polaritonic properties (group velocity, quality factor of localized modes, emission pattern, etc.) and can be applied to a wide range of excitonic materials. As a proof-of-concept, the strong coupling regime at room temperature in perovskite-based metasurfaces is experimentally observed. Moreover, we demonstrate the dispersion engineering with a variety of polariton dispersion characteristics such as linear, slow light, and for the first time multi-valley polaritonic mode dispersion, respectively. Finally, our approach is naturally in the scheme of integrated optics,  and perfectly adapted for large-scale fabrication methods such as nanoimprint and solution spincoating. It thus would pave the way for making low-cost integrated polaritonic devices operating at room temperature.\\

\subsection{Methods}

\label{Methods}

\textbf{Fabrication of the patterned SiO$_2$ backbone:} The negative pattern (i.e. hole lattices) is defined in PMMA A4 950 Microchem (resist) using electron-beam lithography  system equipped with Raith Elphy pattern generator. Before exposure, a film of 10\,nm Al is deposited on top of the photoresist via thermal evaporation to prevent surface charging effects. After exposure, the Al film is removed by chemical etch applying aluminum etchant type D solution. The sample is then developed using MIBK/IPA solution. Later, the pattern is transferred  into the fused silica substrate by Reactive Ion Etching using CHF$_3$ gas under pressure of 50\,mTorr. \\

\noindent \textbf{Infiltration and crystallization of perovskite:} In preparation for perovskite deposition, the SiO$_2$ backbone has been through cleaning processes including cleaning with acetone, ethanol in ultrasonic bath and ozone treatment. The treatments play an important role for quality enhancement of thin film deposition.  PEPI solution in DMF wt 10\% is spincoated on top of pre-treated SiO$_2$ backbone at 2000-3000\,rpm within 30\,s. The crystallization of PEPI pillars is achieved by annealing at $95^o$C during 90\,s. At last, a thin film of 200\,nm of PMMA (resist) is deposited on top of PEPI metasurface by spincoating at 3000\,rpm in 30\,s and followed by an annealing process at $95^o$C during 15\,min. \\

\noindent \textbf{ARR and ARPL setup:} The excitation sources for ARR and ARPL are a halogen light and a picosecond pulsed laser (50\,ps, 80\,MHz, 405\,nm) respectively. The excitation light is focused onto the sample via a microscope objective (x100, NA~=~0.8), The excitation spot-size is within $1\,\mu$m for the ARPL and 5\,$\mu$m for the ARR measurements. The emitted light is collected via the same objective and its image in the Fourier space is projected onto the entrance slit of a spectrometer. The sample orientation is aligned so that the $\Gamma X$ direction is along the entrance slit. The output of the spectrometer is coupled to a CCD camera, and the image obtained from the camera leads directly to the energy-momentum dispersion diagram along $\Gamma X$. \\

\noindent \textbf{PEPI model:} The dielectric function of PEPI, used in numerical simulations and analyical calculations, is given by:
\begin{align}
    \epsilon_{PEPI}(E)=n^2+\frac{A_X}{E_X^2-E^2-i\gamma_XE}
\end{align}
where $n=2.4$ is the refractive index of the passive structure, $A_X=0.85\,eV^2$ is the oscillator strength of PEPI exciton, $E_X=2.394\,eV$ is its energy and $\gamma_X=30\,meV$ is its linewidth~\cite{Zhang2010}. \\

\noindent\textbf{Numerical simulations:} The RCWA simulations of angular-resolved reflectivity and absorption have been done with S\textsuperscript{4}, a freely available software package provided by Fan Group in the Stanford Electrical Engineering Department~\cite{Liu2012}. \\

\noindent\textbf{Quantum theory of exciton-photon coupling:} The radiation-matter coupling in periodically patterned multilayers is described by quantizing both the electromagnetic field and the exciton center of mass field in a periodic piecewise constant potential. Photonic modes, in particular, are obtained through a guided mode expansion method \cite{Andreani2006}, which allows to obtain both real and imaginary parts (losses) of photonic Bloch modes of the textured structure. The second-quantized hamiltonian of the interacting system is then  diagonalized with a generalized Hopfield method, thus yielding the complex dispersion of mixed exciton-photon modes \cite{Gerace2007}. The main approximation with respect to the present work lies in the assumption that the whole excitonic oscillator strength is concentrated at the center of the patterned layer, while the full thickness of the HOP layer is taken into account with a non-dispersive average index (2.4). This way, the only `fitting' parameter with respect to experimental results is the effective oscillator strength assumed for this single quantum well layer. On the other hand, we assumed a value for the oscillator strength per unit surface ($f/S$) that is actually obtained by multiplying the oscillator strength per quantum well in the HOP layer times the number of wells compatible with the overall HOP layer thickness. So, without adjustable parameters, our assumption proves to be reliable {\it a posteriori}, by the remarkable agreement with experimental results for the polariton dispersions reported in Fig.~2 and 3 of this manuscript. \\

\begin{acknowledgement}

The authors would like to thank the staff from the NanoLyon Technical Platform for helping and supporting in all nanofabrication processes. We thank Jos\'e Penuelas for his technical support in the XRD characterization of perovskite. This work is partly supported by the French National Research Agency (ANR) under the project POPEYE (ANR-17-CE24-0020) and project EMIPERO (ANR-18-CE24-0016).

\end{acknowledgement}




\section*{Author information}

\subsection{Corresponding Authors}
$*$Email: hai-son.nguyen@ec-lyon.fr
\subsection{Author Contributions}
C.S., R.M., and N.H.M.D. implemented the fabrication of nano-patterned subtrates. G.T.-~A., F.L., and E.D. worked on PEPI perovskite synthesis and its excitonic model. E.D., H.S.N. and N.H.M.D carried out numerical simulation. D.G. developed the quantum model of \textit{photonic crystal} polaritons. H.S.N. and N.H.M.D worked on perovskite deposition and performed the optical experiments. All the authors contributed to the interpretation of the results.


\subsection{Notes}
The authors declare no competing financial interest.

\bibliography{mybib}


\newpage
\section{Figures and Captions}
\begin{figure}
    \centering
    \includegraphics[width=1\textwidth]{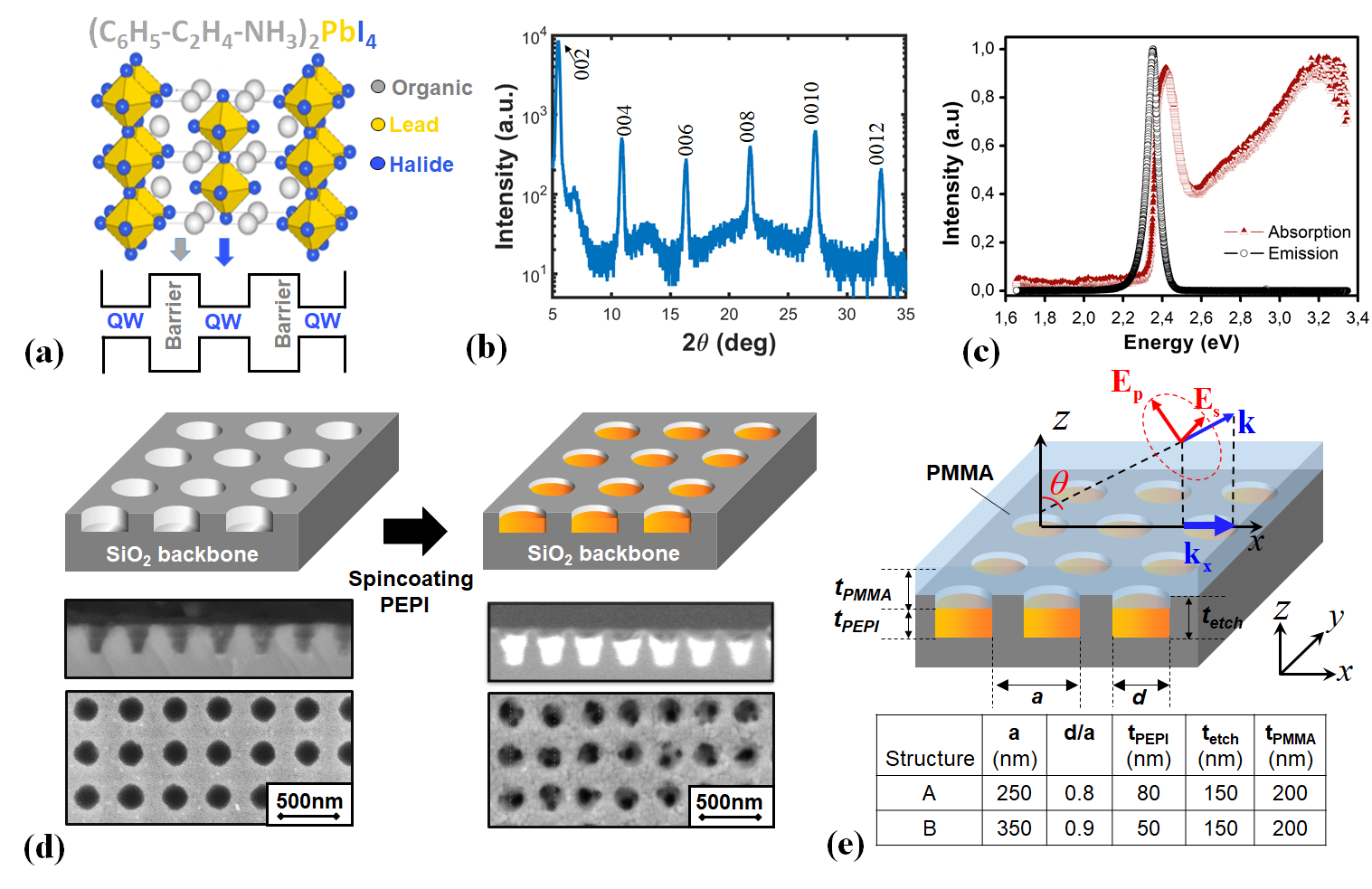}
    \caption{(a) Molecular structure of PEPI. (b) XRD spectrum of 50\,nm PEPI on glass substrate (with PMMA coated on top). (c) Absorption and photoluminescence spectrum of thin film PEPI, excited at wavelength 405\,nm. (d) Fabrication scheme and SEM charaterization results of PEPI metasurface. (e) Sample design and demonstration of characterization}
    \label{fig1}
\end{figure}

\begin{figure}
    \centering
    \includegraphics[width=1\textwidth]{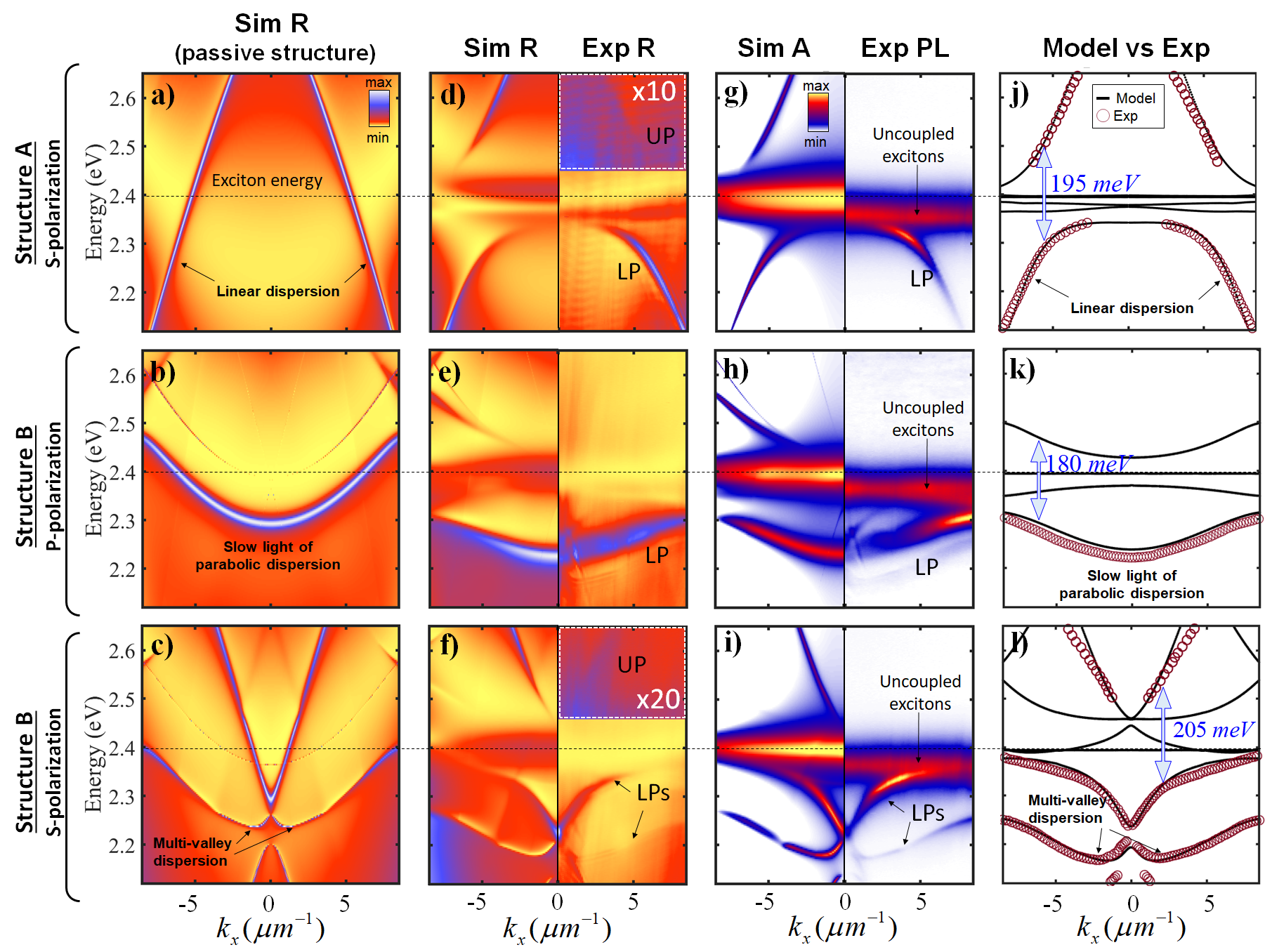}
    \caption{(a,b,c) Numerical simulations of the ARR for the passive structures by a RCWA code. (d,e,f) Experimental results (right panels) and numerical simulations (left panels) of the ARR spectra for the active structures. (g,h,i) Experimental results of the ARPL response (right panels) compared to numerical simulations of the angular-resolved absorption by RCWA (left panels) for the active structures. (j,k,l)  Numerical solution of photonic crystal polariton dispersions from a quantum model of radiation-matter interaction, compared to experimental results extracted from ARR measurements. In all figures, the black dash line represents the energy of the PEPI excitonic resonance.}
    \label{fig2}
\end{figure}

\begin{figure}
\begin{center}
\includegraphics[width=0.5\textwidth]{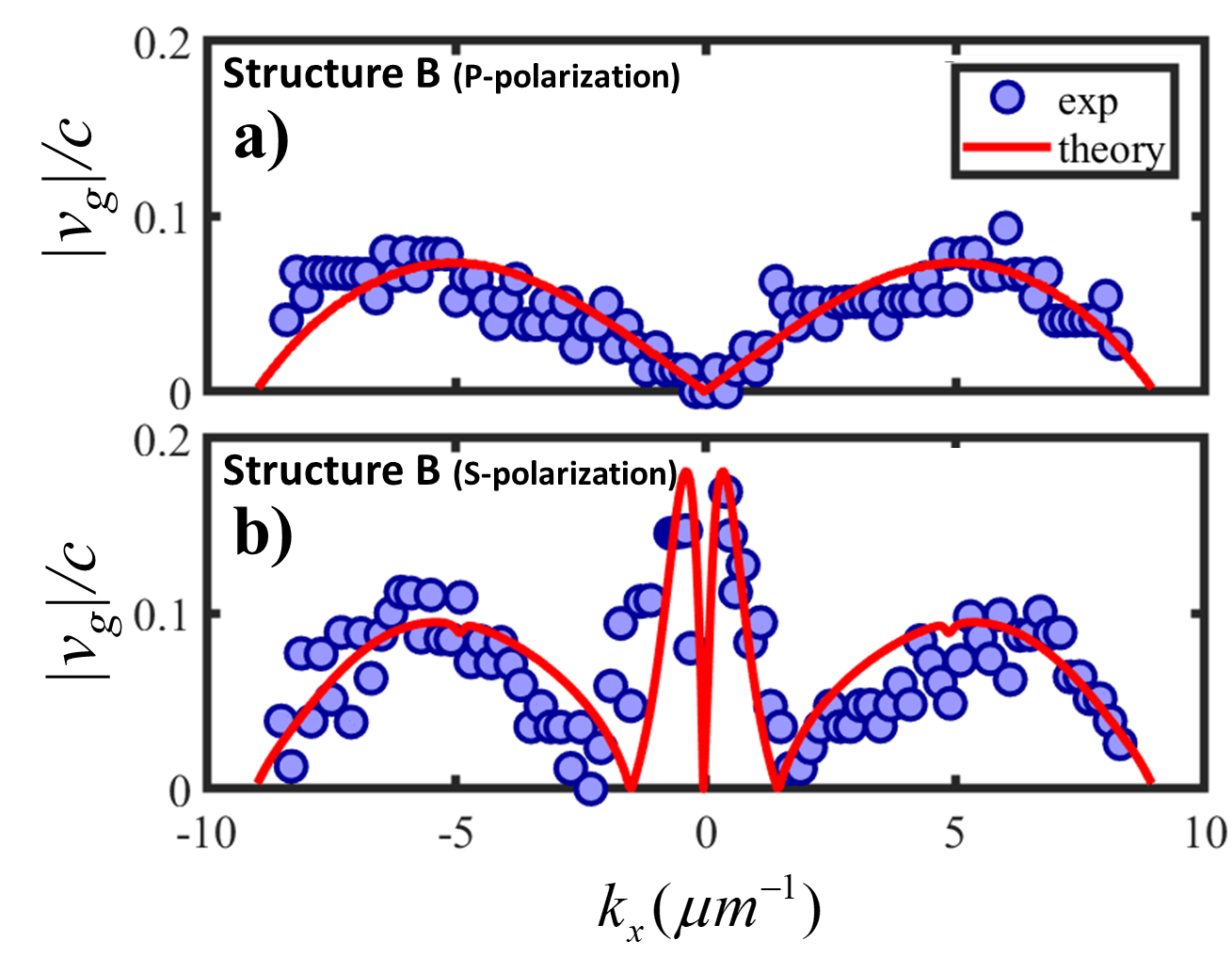}
\end{center}
\caption{Extracted group velocity  (absolute value normalized by the speed of light) of (a) P-polarized LP in structure B, (b) S-polarized LP of multi-valley dispersion in structure B. The solid red lines correspond to analytical calculations. The light-blue circles are experimental data extracted from Fig~2(k,l).}
\label{fig3}
\end{figure} 

\newpage
\begin{tocentry}
    \includegraphics[width=1\textwidth]{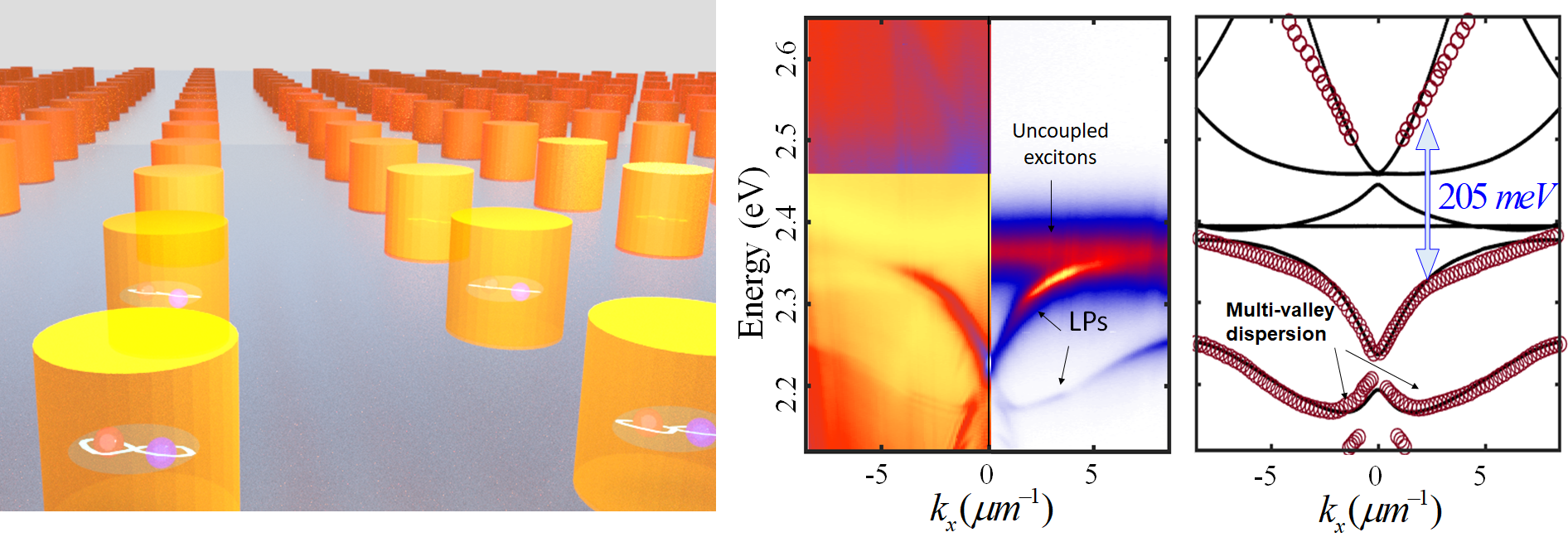} 
\end{tocentry}
\end{document}